\documentclass[12pt]{iopart}

\usepackage{iopams}
\usepackage{amssymb}
\usepackage{graphicx}

\begin{document}

\title[Correlations between Maxwell's multipoles]{Correlations between Maxwell's multipoles for gaussian random functions on the sphere}

\author{M R Dennis\footnote{Present address: School of Mathematics, University of Southampton, Highfield, Southampton SO17 1BJ, UK}}

\address{H H Wills Physics Laboratory, Tyndall Avenue, Bristol BS8 1TL, UK}

\begin{abstract}
Maxwell's multipoles are a natural geometric characterisation of real functions on the sphere (with fixed $\ell$). The correlations between multipoles for gaussian random functions are calculated, by mapping the spherical functions to random polynomials. In the limit of high $\ell,$ the 2-point function tends to a form previously derived by Hannay in the analogous problem for the Majorana sphere. The application to the cosmic microwave background (CMB) is discussed.
\end{abstract}

\pacs{02.30.Px, 05.45.Mt, 98.70.Vc}

\begin{figure}
\begin{center}
\includegraphics*[width=6cm]{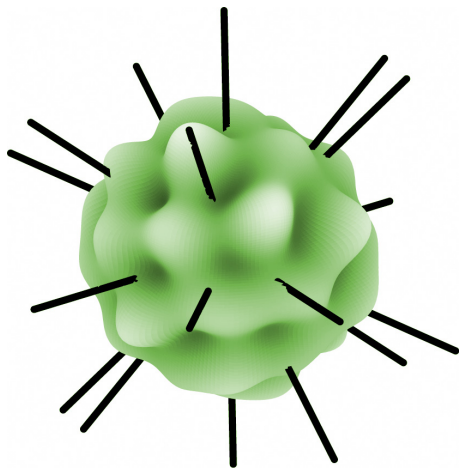}
\end{center}
    \caption{A sample spherical function $\Phi$ for $\ell = 10.$ The radius of the plot is determined by $\Phi(\theta,\phi)$ plus a constant, and shaded according to the value of $\Phi.$ The Maxwell multipole directions for the function are also represented.}
    \label{fig:polesex}
\end{figure}

A striking feature of randomness is the emergence of structure arising from independent processes. 
Here, I describe a simple, universal correlation structure associated with statistically isotropic random functions on the sphere: the correlations between Maxwell's multipoles.
In the multipole representation, any real spherical function $\Phi(\theta,\phi),$ with fixed $\ell,$ may be represented by $\ell$ unit vectors $\bi{u}_1, \dots, \bi{u}_{\ell},$
\begin{equation}
\Phi(\theta, \phi) = \sum_{m = -\ell}^{\ell} a_m Y_{\ell}^m(\theta, \phi) = C (\bi{u}_1\cdot \nabla)\cdots(\bi{u}_{\ell}\cdot \nabla)\frac{1}{\sqrt{x^2 + y^2 + z^2}}
\label{eq:spherrep}
\end{equation}
for $a_{-m}=(-1)^m a_m^{\ast}$ (ensuring reality of $\Phi$), $(x, y, z)$ restricted to the sphere and $C$ a numerical constant.
The $\ell$ vectors $\bi{u}_i$ are called Maxwell's multipoles \cite{maxwell:treatise1}, and are determined uniquely (up to sign) \cite{sylvester:note, ch:methods1, dennis:canonical}.
The significance of the Maxwell representation is that the multipoles rotate directly with the function, and are defined without reference to an external reference frame (unlike spherical harmonics).
An arbitrary spherical function with fixed $\ell,$ with its corresponding multipole directions, is shown in figure \ref{fig:polesex}.

The quantity discussed in this note is the correlation of the Maxwell multipole directions for completely random functions of the form (\ref{eq:spherrep}). 
Complete randomness means that the probability density function of $\Phi$ depends only on the sum $\sum_{m = 0}^{\ell} |a_m|^2;$ the statistical distribution of the multipoles (which is automatically rotationally symmetric) is determined absolutely by this restriction.
Here a gaussian distribution of the sum is chosen for calculations, which implies that the $a_m$ coefficients can be taken as independent (for $m \ge 0$), identically distributed gaussians (by analogy with the derivation of the Maxwell distribution in kinetic theory).

The statistical morphology of cosmic microwave background (CMB) radiation is still little understood, and it has been suggested that correlations between Maxwell multipoles might reveal hidden structure in the data \cite{chs:multipole, sshc:lowl, lm:multipole}.
Here, I will derive the universal correlation function of Maxwell multipoles for a statistically isotropic spherical function, using independent gaussian random $a_{m}$ coefficients.
The present analytic approach contrasts with the numerical Monte Carlo computations for the same isotropic gaussian distribution studied in the papers referred to above.

The main tool used in the calculation is the Majorana representation of spherical functions (also called the `stellar representation') \cite{dennis:canonical, majorana:atomi, bacry:orbits, penrose:road, hannay:zero}. 
In this representation, the function $\Phi(\theta,\phi)$ with the form (\ref{eq:spherrep}) is represented as a polynomial in the complex variable $\zeta,$
\begin{equation}
   f(\zeta) = \sum_{m = -\ell}^{\ell} (-1)^m \left(\begin{array}{c} 2\ell \\ \ell+m \end{array} \right)^{1/2} a_m \zeta^{\ell + m}.
   \label{eq:majorana}
\end{equation}
The relationship with the sphere is realised by taking $\zeta = \exp(\rmi\phi)\tan(\theta/2)$ as a stereographic coordinate in the complex plane.
Since each $a_{-m}=(-1)^m a_m^{\ast},$ the roots of the polynomial $f$ occur in antipodal pairs $\zeta_i, -1/\zeta_i^{\ast};$ these are the stereographic projections of the Maxwell multipole vectors (whose sign is undefined) \cite{dennis:canonical}.
For complex functions on the sphere (with independent $a_m$), spinor-like Majorana directions arise from the roots of (\ref{eq:majorana}). 
(In the language of group theory, basis functions of the unitary (rotation) group SU(2) are mapped to those of the linear group SL(2,$\mathbb{C}$) in this representation.)

The identification of the spherical function (\ref{eq:spherrep}), with gaussian random coefficients, with the polynomial (\ref{eq:majorana}), reduces the problem to a calculation of the statistics of roots of random polynomials.
The roots of spin-like random polynomials (SU(2) polynomials) have been studied, particularly in connection with quantum chaotic systems \cite{hannay:zero, bbl:distribution,  prosen:exact, bbl:quantum, bkz:timereversal, hannay:caf, leboeuf:random, fh:exact, bsz:universality}, with the distribution (\ref{eq:majorana}), but different restrictions on the identically distributed gaussian coefficients $a_m.$

The present calculation is similar to that of Hannay \cite{hannay:zero}, who derived the $k$-point correlation function $\rho_k$ between the Majorana spinor directions for complex functions analogous to $\Phi$ in equation (\ref{eq:spherrep}), that is, with no additional requirements on the coefficients $a_m.$
The 2-point function $\rho_2,$ in the large $\ell$ limit, was found to have a simple universal form.
Following the same method, Maxwell's multipoles are found here to limit to the same function.
Readers not interested in the details of the calculation may skip to equation (\ref{eq:rho21}) and the following discussion.

As stated above, $\Phi$ in equation (\ref{eq:spherrep}) is assumed to have coefficients $a_m$ independent and identically gaussian distributed (which implies, but is not implied by, isotropy):
\begin{equation}
   \langle a_m^{\ast} a_n \rangle = \delta_{m,n}, \quad \langle a_m a_n \rangle = (-1)^m \delta_{m,-n}.
   \label{eq:iid}
\end{equation}
All statistical information about the multipoles is in the zeros of the Majorana polynomial (\ref{eq:majorana}), which are accessed by averaging $f_i \equiv f(\zeta_i)$ at the point $\zeta_i,$ and its derivative $f'_i\equiv \rmd f(\zeta)/\rmd \zeta |_{\zeta = \zeta_i}.$
The joint probability density function $P$ for $k$ points $f_1, \dots, f_k, f^{\ast}_1, \dots, f^{\ast}_k, f'_1, \dots, f'_k, f'^{\ast}_1, \dots, f'^{\ast}_k$ is given by the gaussian distribution
\begin{equation}
\fl P(f_1, \dots, f_k, f^{\ast}_1, \dots, f^{\ast}_k, f'_1, \dots, f'_k,
f'^{\ast}_1, \dots, f'^{\ast}_k)
= \frac{\exp(-\bi{F}^{\ast}\cdot \mathbf{M}^{-1} \cdot \bi{F}/2)}{\pi^k
\sqrt{\det\mathbf{M}} },
   \label{eq:pdef}
\end{equation}
where $\bi{F} = (f_1, \dots, f_k, f^{\ast}_1, \dots, f^{\ast}_k, f'_1, \dots, f'_k, f'^{\ast}_1, \dots, f'^{\ast}_k),$ and $\mathbf{M}$ is the $4k\times4k$ hermitian correlation matrix $\mathbf{M}_{ij} = \langle F^{\ast}_i F_j \rangle.$ 
Following \cite{hannay:zero}, submatrices (of dimension $2k$) of $\mathbf{M}$ will be identified:
\begin{equation}
   \mathbf{M} = \left( \begin{array}{cc} \mathbf{A} & \mathbf{B} \\ \mathbf{B}^{\dagger} & \mathbf{C} \end{array} \right).
   \label{eq:mdecomp}
\end{equation}

Nonvanishing terms (and their conjugates) appearing in $\mathbf{M}$ are \cite{hannay:zero}
\begin{eqnarray}
\fl   \langle f_i^{\ast} f_j \rangle 
   = (1 + \zeta_i^{\ast} \zeta_j)^{2\ell}, &\quad&
   \langle f_i f_j \rangle 
   = (\zeta_i - \zeta_j)^{2\ell},
   \nonumber\\
\fl   \langle f_i'^{\ast} f_j \rangle 
   = 2\ell \zeta_j (1 + \zeta_i^{\ast} \zeta_j)^{2\ell-1},  &\quad&
   \langle f_i' f_j \rangle 
   = 2\ell (\zeta_i - \zeta_j)^{2\ell-1},
   \nonumber \\
\fl   \langle f_i^{\ast} f'_j \rangle 
   = 2\ell \zeta_i^{\ast} (1 + \zeta_i^{\ast} \zeta_j)^{2\ell-1}, &\quad&
   \langle f_i f_j' \rangle 
   = -2\ell (\zeta_i - \zeta_j)^{2\ell-1},
   \nonumber \\
\fl   \langle f'^{\ast}_i f'_j \rangle 
   = 2\ell (1+2\ell \zeta_i^{\ast} \zeta_j) (1 + \zeta_i^{\ast} \zeta_j)^{2\ell-2}, &\quad&
   \langle f_i' f_j' \rangle 
   = -2\ell(2\ell-1) (\zeta_i - \zeta_j)^{2\ell-2},
   \label{eq:corrvals}
\end{eqnarray}
following from averaging the complex polynomials (\ref{eq:majorana}) using (\ref{eq:iid}).

The $k$-point function $\rho_k(\zeta_1, \dots, \zeta_k)$ is found using standard methods for finding the correlations of zeros of gaussian random functions (e.g. \cite{dennis:correlations}):
\begin{eqnarray}
\fl   \rho_k(\zeta_1, \dots, \zeta_k) 
   &=& \langle \delta(f_1)\delta(f_1^{\ast}) \cdots \delta(f_k)\delta(f_k^{\ast}) |f_1' \cdots f_k'|^2 \rangle 
   \nonumber \\
   &=& \frac{1}{\pi^k \sqrt{\det{\mathbf{M}}}}\int \rmd^{4k}\bi{F} \delta(f_1)\cdots \delta(f_k^{\ast}) |f_1' \cdots f_k'|^2 \exp(-\bi{F}^{\ast}\cdot \mathbf{M}^{-1} \cdot \bi{F}/2)
   \nonumber \\
   &=& \frac{1}{\pi^k \sqrt{\det{\mathbf{M}}}}\int \rmd^{2k}\bi{F}' |f_1' \cdots f_k'|^2 \exp(-\bi{F}'^{\ast}\cdot \mathbf{N}^{-1} \cdot \bi{F}'/2),
   \label{eq:rhok}
\end{eqnarray}
where $\bi{F}' = (f_1', \dots, f_k',f_1'^{\ast},\dots f_k'^{\ast}),$ and $\mathbf{N}^{-1}$ is the submatrix of $\mathbf{M}^{-1}$ in the position of $\mathbf{C}$ in $\mathbf{M}.$
By Jacobi's determinant theorem \cite{hannay:zero, dennis:correlations}, $\mathbf{N} = \mathbf{C} - \mathbf{B}^{\dagger} \mathbf{A}^{-1} \mathbf{B},$ and $\det\mathbf{M} = \det\mathbf{A} \det\mathbf{N}.$

The final part of the calculation is a gaussian integration by parts (i.e. the finite dimensional analogue of Wick's theorem), and the final result is $(\pi^k \sqrt{\det{\mathbf{A}}})^{-1}$ times a combinatorial term.
This term is the sum over products of elements of $\mathbf{N},$ one for each pairing of the $2k$ terms $f'_1, \dots f_k'^{\ast}.$
The elements of $\mathbf{N}$ in the product are those which correspond to the appropriate components of $\bi{F}', \bi{F}'^{\ast}$ in the quadratic form.

The simplest nontrivial case is the 2-point function $\rho_2,$ which is
\begin{equation}
   \rho_2 = (N_{11} N_{22} + N_{12} N_{21} + N_{14} N_{41})/(\pi^2 \sqrt{\det{\mathbf{A}}})
   \label{eq:rho2}
\end{equation}
It should be noted that, since the $f_i$ are correlated with each other (rather than simply with their conjugates), this combinatorial term is rather more complicated than the simple permanent found in \cite{hannay:zero}; furthermore, the choice of some elements of $\mathbf{N}$ is not unique (for example, in the third summand in the numerator in (\ref{eq:rho2}), $N_{14} = N_{23}, N_{41} = N_{32}$).

\begin{figure}
\begin{center}
\includegraphics*[width=10cm]{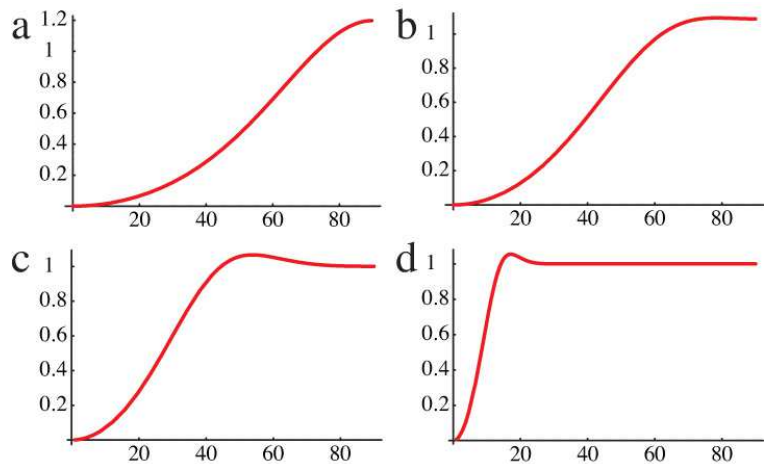}
\end{center}
    \caption{The 2-point correlation function on the sphere $\rho_{\rm{sphere}},$ plotted against $\theta$ (in degrees) for multipoles with separation $\theta$ for (a) $\ell = 3;$ (b) $\ell = 5;$ (c) $\ell = 10;$ (d) $\ell = 100.$ The function has been divided by the square of multipole density $\ell/2\pi$ (so approaches 1 for large $\ell$), and the behaviour for small  $\theta$ is quadratic.}
    \label{fig:rho2}
\end{figure}

The 1-point function $\rho_1$ is simply the density of roots in the complex plane (counting each multipole direction, undetermined in sign, twice).
On the sphere, this is uniform with value $2\ell/4\pi.$
The 2-point function $\rho_2$ can be simplified by taking the two points to be $0$ and $r$ (with $r = \tan(\theta/2)$ for points with angular separation $\theta$); no generality is lost since the distribution is rotationally symmetric on the sphere.
From (\ref{eq:rho2}), this is
\begin{eqnarray}
\fl   \rho_2(0,r) = (\pi^2 D^{5/2})^{-1} \{ 
   [2\ell D - 4 b u v - (b^2 + v^2)(a - 1 - u^2) ]  \nonumber \\
   \times [d D - 2 c u v (a + 1 - u^2) - (c^2+ a v^2)(a - 1 - u^2) ]\nonumber \\
   + [2\ell D - 2 c u v - b u v (a + 1 -u^2) - v^2 (a - 1 + u^2) - b c (a - 1 - u^2)]^2   \nonumber \\
    + [w D - 2 b c u - u v^2 (a + 1 - u^2) - b v (a - 1 + u^2) - c v (a - 1 - u^2)]^2 \}
\label{eq:rho21}
\end{eqnarray}
with $D = \det \mathbf{A} = (a-1-u^2-2u)(a-1-u^2+2u)$ and, extending the notation of \cite{hannay:zero},
\begin{eqnarray}
\fl  a = (1 + r^2)^{2\ell}, \, b = 2\ell r, \, c = 2\ell r (1 + r^2)^{2\ell-1}, \,  d = 2\ell (1+2\ell r^2)(1 + r^2)^{2\ell-2}, \nonumber \\
   u = r^{2\ell}, \, v = -2\ell r^{2\ell-1}, \, w = -2\ell(2\ell-1) r^{2\ell-2}.
   \label{eq:corrvals1}
\end{eqnarray}
To find the 2-point multipole correlation function function on the sphere $\rho_{\rm{sphere}}(\theta),$ for points with angular separation $\theta,$  this must be multiplied by the stereographic jacobian $(1+r^2)^2/16$ and substituting $r = \tan(\theta/2)$ \cite{hannay:zero}.
$\rho_{\mathrm{sphere}}(\theta)$ is symmetric about $\pi/2,$ by antipodality of the multipole directions.
The function may be normalised (i.e. is unity when points are uncorrelated) by dividing by the square of the density $2\ell/4\pi.$
When $\ell = 2,$ $\rho_{\rm{sphere}}(\theta)$ can be found from equation (\ref{eq:rho21}),
\begin{equation}
   \rho_{\rm{sphere}}(\theta) = \left(\frac{4}{4\pi}\right)^2\frac{27(1-\cos^2 \theta)}{2(3+\cos^2\theta)^{5/2}} \qquad \qquad \hbox{(for $\ell = 2$)},
   \label{eq:rhosphere}
\end{equation}
agreeing up to a numerical constant with that found by \cite{lm:multipole}.
$\rho_{\rm{sphere}}$ is plotted against $\theta$ in figure \ref{fig:rho2} for some choices of $\ell.$
There is good agreement with the corresponding numerical results of \cite{lm:multipole} (who plot against $\cos \theta$).

In \cite{hannay:zero}, it was found, for a general complex isotropic Majorana polynomial, in the limit of large $\ell,$ that $\rho_2(0,r)$ tends to $(\ell/2\pi)^2 g(\sqrt{\ell} r),$ where
\begin{equation}
   g(R) = \left((\sinh^2 R^2+R^4)\cosh R^2 - 2 R^2 \sinh R^2\right)/\sinh^3 R^2.
   \label{eq:hannayg}
\end{equation}
In equation (\ref{eq:rho21}), on writing $r = R/\sqrt{\ell}$ and taking the large $\ell$ limit (in which $u, v,$ and $w$ approach $0$), equation (\ref{eq:rho21}) factorizes to give the 2-point function of this general complex case (equation (11) in \cite{hannay:zero}).
In the complex plane, this large $\ell$ limit of the general random Majorana polynomials is the universal `chaotic analytic function' \cite{hannay:caf, bsz:universality}.

Different symmetries for the random polynomial coefficients have been investigated, such as for $a_m$ real \cite{prosen:exact}.
In this case, the distribution of roots in the complex plane behaves singularly on the real axis (where roots appear in complex conjugate pairs), and approaches the chaotic analytic function behaviour far from these lines (where there is little influence from the conjugate, paired zero).
(The generalization of random polynomials with coefficients of fixed argument \cite{bkz:timereversal} have similar behaviour.)
The distribution of roots of the Maxwell multipole polynomials is uniform on the (stereographic) sphere, and the difference with the chaotic analytic function polynomials is the presence of the antipodal root, whose influence diminishes as $\ell$ increases.
Therefore, for high $\ell,$ the correlations between Maxwell multipoles also limit to the form (\ref{eq:hannayg}), apparently by a power law.

$g(R)$ exhibits quadratic repulsion at the origin with a small maximum of $1.0531$ at $R = 1.4985$, and thereafter approaches 1 exponentially.
For high $\ell,$ the maximum of $\rho_2$ (visible in figure \ref{fig:rho2}) is approximately at $\theta = 3/2\sqrt{\ell}.$
In this limit, the 2-point function of Maxwell multipoles is therefore constant (there is no correlation) apart from a tiny repulsion region near each multipole.
This is analogous to the closely related system of a one-component plasma on the sphere \cite{caillol:exact}, whose 2-point correlation function, in the thermodynamic limit, approaches a constant apart from repulsion very near the origin.

Higher correlation functions between more Maxwell multipole directions, which are universal for completely random spherical functions, can also be calculated using the methods described here.
For large $\ell,$ these will tend to the same behaviour as high $\ell$ random Majorana polynomials \cite{hannay:zero, bsz:universality}.
However, it is possible that lower $\ell$ correlations might reveal more subtle structure, which may be compared with CMB data.

\ack
I am grateful to Mark Birkinshaw and John Hannay for discussions, and to Jeff Weeks who pointed out to me the relevance of Maxwell's multipoles to CMB. This work was supported by the Leverhulme Trust.

\end{document}